\documentstyle[sprocl,epsfig]{article}

\def\Journal#1#2#3#4{{#1} {\bf #2}, #3 (#4)}

\def\apj{\em Astrophys. J.}
\def\apjs{\em Astrophys. J. Suppl.}
\def\euro{\em Europhys. Lett.}
\def\NCA{\em Nuovo Cimento}

\def\NPB{{\em Nucl. Phys.} B}
\def\npps{\em Nucl. Phys. (Proc. Suppl.)}
\def\PLB{{\em Phys. Lett.}  B}
\def\ppnp{\em Prog. in Part. and Nucl. Phys.}
\def\PRL{\em Phys. Rev. Lett.}
\def\PRC{{\em Phys. Rev.} C}
\def\PRD{{\em Phys. Rev.} D}
\def\Science{\em Science}
\def\sjnp{\em Sov. J. Nucl. Phys.}
\def\ZPC{{\em Z. Phys.} C}

\def\be{\begin{equation}}
\def\ee{\end{equation}}
\def\bea{\begin{eqnarray}}
\def\eea{\end{eqnarray}}
\def\gtwid{\mathrel{\raise.3ex\hbox{$>$\kern-.75em\lower1ex\hbox{$\sim$}}}}
\def\ltwid{\mathrel{\raise.3ex\hbox{$<$\kern-.75em\lower1ex\hbox{$\sim$}}}}

\begin{document}

\title{NEUTRINO MASS AND DARK MATTER}

\author{ DAVID O. CALDWELL~\footnote{Supported in part by the
 U.S.~Department of Energy.} }

\address{Institute for Nuclear and Particle Astrophysics and Cosmology
 and\\ Physics Department, University of California, Santa Barbara, CA
 93106-9530, USA}


\maketitle\abstracts{
Despite direct observations favoring a low mass density, a critical density
universe with a neutrino component of dark matter provides the best existing
model to explain the observed structure of the universe over more than three
orders of magnitude in distance scale.  In principle this hot dark matter
could consist of one, two, or three species of active neutrinos.  If all
present indications for neutrino mass are correct, however, only the
two-species ($\nu_\mu$ and $\nu_\tau$) possibility works.  This requires
the existence of at least one light sterile neutrino to explain the solar
$\nu_e$ deficit via $\nu_e\to\nu_s$, leaving $\nu_\mu\to\nu_\tau$ as the
explanation for the anomalous $\nu_\mu/\nu_e$ ratio produced by atmospheric
neutrinos, and having the LSND experiment demonstrating via $\bar\nu_\mu\to
\bar\nu_e$ the mass difference between the light $\nu_e$--$\nu_s$ pair and
the heavier $\nu_\mu$--$\nu_\tau$ pair required for dark matter.  Other
experiments do not conflict with the LSND results when all the experiments
are analyzed in the same way, and when analyzed conservatively the LSND
data is quite compatible with the mass difference needed for dark matter.
Further support for this mass pattern is provided by the need for a sterile
neutrino to rescue heavy-element nucleosynthesis in supernovae, and it
could even aid the concordance in light element abundances from the early
universe.}

\section{Introduction}

Just when the issue of neutrino mass is clarifying, the cosmological
consequences of such mass has become more puzzling.  About 100/cm$^3$
of relic neutrinos are everywhere, and the evidence for neutrino mass
is now generally accepted.  It is even likely that there are three neutrino
mass differences, the basis for which will be reviewed briefly, but
experiments have not yet proved that the mass is sufficient to have
important cosmological effects.  On the basis not of experiments but of
observations, the situation is remarkably confusing.  On the one hand,
there are many observations showing the matter density of the universe
($\Omega_m$) to be less than critical ($\Omega_m<1$).  On the other hand,
the only model which fits observations of universe structure over more
than three orders of magnitude in distance scale is one having cold plus
hot (presumably neutrinos) dark matter and $\Omega_m=1$.  These observations
exclude models of an open universe (low $\Omega_m$) or one adding to cold
dark matter a cosmological constant ($\Lambda$) so as to make critical density
($\Omega_m+\Omega_\Lambda=1$).  Adding hot dark matter helps only a little
if $\Omega_m\ll1$.

Either some of these observations are wrong, or are being wrongly interpreted,
or something entirely new is needed.  While much more complete and precise
observations will be available soon, the definitive answer regarding the
hot dark matter component provided by neutrinos must come from terrestrial
experiments.  The only experiment providing possible direct evidence for
neutrino mass which could be cosmologically significant is that of LSND,
and its results are interpreted here in a way which enhances that possibility.
The experiments on the solar $\nu_e$ deficiency and the anomalous
$\nu_\mu/\nu_e$ ratio from atmospheric neutrinos are important in this context
in establishing the pattern of neutrino masses to learn which neutrinos
could contribute to the hot dark matter.  The pattern which emerges is made
more likely by indirect evidence from the need to rescue heavy-element
production by supernovae and possibly by removing a small lack of concordance
in the initial abundance of light elements.

\section{Neutrino Dark Matter}

There is now evidence from the first Doppler peak observed in the cosmic
microwave background radiation that the total energy density of the universe
is the critical value; i.e., $\Omega=1$, and the universe will expand forever
at an ever decreasing rate.  Such a flat universe has the only time-stable
value of density and is expected in all but very contrived models of an
early era of exponential expansion, or ``inflation''.  It has usually been
assumed that $\Omega=\Omega_m$; that is, the energy density is the matter
density.  Recent evidence points to $0.3\le\Omega_m\le0.6$, however, based
on a variety of observations: high-redshift supernovae type Ia, evolution of
galactic clusters, high baryon content of clusters, lensing arcs in clusters,
and dynamical estimates from infrared galaxy surveys.  On this basis it has
become popular to assume that $\Omega_m\approx0.3$, but $\Omega=1$ through the
addition of a vacuum energy density, usually designated as a cosmological
constant, $\Lambda$.  In stark contrast to this information, either a
low-density universe or a critical density universe with a cosmological
constant certainly does not fit universe structure as measured over three
orders of magnitude in distance scale by the cosmic microwave background
and galaxy surveys.  The only model (CHDM) which fits these extensive data is
one having $\Omega_m=1$, of which $\Omega_\nu=0.2$ is in neutrinos, and
$\Omega_b=0.1$ in baryons, with the main component being cold dark matter.
This is work of Gawiser and Silk,\cite{ref:1} who used all published data
from the cosmic microwave background and galaxy surveys.  They compared the
data with ten models of universe structure, but of concern here are only
three of these, CHDM, an open universe model (OCDM) having $\Omega_m=0.5$,
and one ($\Lambda$CDM) having $\Omega_m=0.5$ and $\Omega_\Lambda=0.5$.  In
the latter two cases the parameters were varied to get the best fits, resulting
in $\Omega_m=0.5$, of which $\Omega_b=0.05$ with the rest as cold dark matter.
The probabilities of the fits were $\rm CHDM=0.09$,
$\rm OCDM=2.9\times10^{-5}$, and $\rm\Lambda CDM=1.1\times10^{-5}$.  If one
dubious set of data is removed, the APM cluster survey (which disagrees with
galaxy power spectra), these probabilities become $\rm CHDM=0.34$,
$\rm OCDM=6.7\times10^{-4}$, and $\rm\Lambda CDM=4.3\times10^{-4}$.

Had it been possible to extend the fit to even smaller scales, the discrepancy
between CHDM and the others would have been even greater, but this is the
non-linear regime requiring simulations.  The CHDM model with two neutrinos
contributing to $\Omega_\nu$
gives an excellent fit\cite{ref:2} to the data at this extended scale,
whereas the others deviate even more strongly than in the linear region.

Since a model having just baryons and cold dark matter gives a very poor fit
(probability $<10^{-7}$), whereas adding a little hot dark matter makes it
work, the hope naturally arises as to whether a $\Lambda$CDM model with
neutrinos added could be the solution.  Unfortunately, Primack and
Gross\cite{ref:3} have found that the improvement is rather limited.  Having
$\Lambda$ produces a peak in the structure power spectrum which is too large
and at too large a distance scale, and hot dark matter does not contribute
much at that era.

Returning now to the model which does work, in principle the needed neutrino
mass for dark matter could come from one,
two, or three neutrinos, but a fourth one would sufficiently alter the
universe expansion rate at the era of nucleosynthesis to spoil the agreement
between calculations and observed abundances of light elements.  There is now
quite good concordance between the $^4$He abundance values\cite{ref:4} and
the primordial D/H ratio,\cite{ref:5} reinstating the three-neutrino limit
which has been in question recently.  There is a possible way\cite{ref:6}
around this, but it appears to be very unlikely.\cite{ref:7}

\section{Review of Evidence for Nonzero Neutrino Mass}
As will be discussed later, observations do help choose among the one-, two-,
and three-neutrino alternatives for dark matter, but the most discriminating
information comes from experiments, which will now be reviewed briefly.

\subsection{Solar Neutrino Deficit}
All solar neutrino experiments observe fewer electron neutrinos than
solar models predict.  In addition, because the three types of experiments
cover different $\nu_e$ energy ranges and hence sample differently the
contributions from the various nuclear processes producing neutrinos, there is
an energy-dependent discrepancy, exemplified by the relationship
between neutrino fluxes from $^7$Be and $^8$B neutrinos as measured in the
three types of experiments.  The SAGE\cite{ref:8} and GALLEX\cite{ref:9}
radiochemical experiments go to the lowest energy and hence measure all of both
fluxes, while the Homestake\cite{ref:10} radiochemical experiment measures
all of the $^8$B spectrum but only part of the $^7$Be flux, and the
Kamiokande\cite{ref:11} and Super-Kamiokande\cite{ref:12} scattering
experiments measure only $^8$B flux.  Results from
all three actually intersect at a negative value of the
$^7$Be flux, yet $^8$B is produced from $\rm^7Be+p\to\/^8B+\gamma$.  This
problem cannot be avoided by one of the experiments being wrong.  Solar models
which drastically change
solar properties do not solve the problem, and these
models are severely constrained by very accurate helioseismology measurements.

A good solution to the solar $\nu_e$ deficit is provided by oscillation into
$\nu_\mu$, $\nu_\tau$, or $\nu_s$, a sterile neutrino.  While this can be a
vacuum oscillation, requiring a mass-squared difference $\Delta
m^2\sim10^{-10}$ eV$^2$ and large mixing between $\nu_e$ and the other
neutrino, more favored is a matter-enhanced MSW\cite{ref:13} type of
oscillation.  For a $\nu_\mu$ or $\nu_\tau$ final state, $\Delta
m^2_{ei}\sim10^{-5}$ eV$^2$ and mixings either
$\sin^22\theta_{ei}\sim6\times10^{-3}$ or $\sim0.6$ are possible, while only
the former is allowed for $\nu_s$. The main change as a result of the new 
Super-Kamiokande data is that the lack of a day-night effect has reduced the
parameter space for the large-angle solution for the $\nu_\mu$ or $\nu_\tau$
final state.\cite{ref:14}

\subsection{Atmospheric Neutrino Anomaly}
Pions produced in the atmosphere would decay via $\pi\to\mu+\nu_\mu,\/\mu\to
e+\nu_\mu+\nu_e$, so that one would expect
$N(\nu_\mu+\bar\nu_\mu)=2N(\nu_e+\bar\nu_e)$, with a small correction for $K$
decays.  The $(\nu_\mu+\bar\nu_\mu)/(\nu_e+\bar\nu_e)$ ratio would be observed
in underground experiments as $\mu^\pm/e^\pm$, and the result is far from the
expected value.  Because the calculated $\mu^\pm$ and $e^\pm$ individual fluxes
are known to $\sim15$\%, whereas much of the uncertainty drops out in the
ratio, the experiments utilize $R=(\mu/e)_{\rm Data}/(\mu/e)_{\rm Calc}$. 
While it
once appeared that there was a discrepancy between water Cherenkov detectors
and tracking calorimeters,\cite{ref:15} the Soudan II
results\cite{ref:16} agree with those from IMB,\cite{ref:17}
Kamiokande,\cite{ref:18} and Super-Kamiokande.\cite{ref:19}

While the statistical evidence for $R$ being less than unity is now quite
compelling, it is the angular distributions of
the $\mu$ and $e$ events which provide the primary evidence that this deviation
of $R$ from unity is explained by neutrino oscillations.  This non-flat
distribution with angle of $R$ was first observed in the high-energy ($>1.3$
GeV) event sample from Kamiokande, but has now been confirmed with better
statistics in the similar data sample from Super-Kamiokande.  The data fit an
oscillation hypothesis, using $\Delta m^2\approx2\times10^{-3}$ eV$^2$,
$\sin^22\theta\approx1$,
and is far from a non-oscillation, flat distribution.  The low-energy
($<1.3$ GeV) sample also agrees with the same oscillation parameters, but this
should be a much shallower angle dependence, and hence it is statistically less
compelling.

The disappearance of the muon neutrinos could be due to $\nu_\mu\to\nu_\tau$ or
$\nu_\mu\to\nu_e$, with $\nu_\mu\to\nu_s$ being unlikely because the large
mixing angle would bring the $\nu_s$ into equilibrium in the early universe,
possibly providing too many neutrinos to get agreement between predictions of
nucleosynthesis and observed light element abundances, as discussed in the
previous section.  The Super-Kamiokande
observations of $e$ and $\mu$ compared to calculated fluxes, as well as the
individual $e$ and $\mu$ angular distributions,
makes $\nu_\mu\to\nu_e$ very unlikely, since the $e$ distributions are like the
non-oscillation Monte Carlo, whereas those for $\mu$ agree with the oscillation
prediction.  The recent results of the CHOOZ nuclear reactor
experiment,\cite{ref:20} which does not see
evidence of $\nu_e$ disappearing in the appropriate region of $\Delta m^2$ and
$\sin^22\theta$, confirms that the atmospheric effect is very unlikely to be 
$\nu_\mu\to\nu_e$.
On the basis that the Super-Kamiokande observed values of $R$ and angular
distributions of $R$ are due to $\nu_\mu\to\nu_\tau$, the likely value of 
$\Delta m^2$ is definitely much larger than that
required for an explanation of the solar neutrino deficit, and the flavors
of neutrinos cannot be the same in the two cases.  Turning now to the third
possible manifestation of neutrino mass, we shall see that the atmospheric
$\Delta m^2$ is much smaller than that required for the LSND experiment, and
hence that three distinctly different values of neutrino mass differences are
required.

\subsection{Evidence from the LSND Experiment}

\begin{figure}[htb]
\centerline{\epsfxsize=7cm
\epsfbox{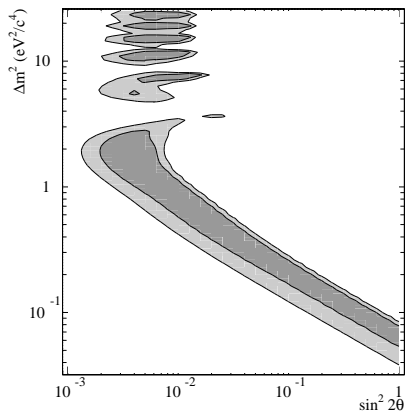}}
\caption{Mass-squared difference ($\Delta m^2$) vs.\ degree of mixing
($\sin^22\theta$) for a $\bar\nu_\mu\to\bar\nu_e$ explanation of the
LSND beam-excess data.  Shown are regions of $\Delta m^2$ favored using
the energy (from 20 to 60 MeV) and distance from the source of each event.}
\label{fig:1}
\end{figure}

The LSND accelerator experiment uses a decay-in-flight $\nu_\mu$ beam of up to
$\sim180$ MeV from $\pi^+\to\mu^+\nu_\mu$ and a decay-at-rest $\bar\nu_\mu$
beam of less than 53 MeV from the subsequent $\mu^+\to e^+\nu_e\bar\nu_\mu$. 
The 1993+1994+1995 data sets included 22 events of the type $\bar\nu_ep\to
e^+n$, expected from $\bar\nu_\mu\to\bar\nu_e$, which was based on
identifying an electron using Cherenkov
and scintillation light that was tightly correlated with a $\gamma$ ($<0.6$\%
accidental rate) from $np\to d\gamma$ (2.2 MeV).  Only $4.6\pm0.6$ such
events were expected from backgrounds.\cite{ref:21}  The chance that these
data, using a water target, result from a fluctuation is $4\times10^{-8}$. 
Note especially that these data were restricted to the energy range 36 to 60
MeV to stay below the $\bar\nu_\mu$ endpoint and to stay above the region where
backgrounds are high due to the $\nu_e\/^{12}{\rm C}\to e^-X$ reaction.  In
plotting $\Delta m^2$ vs.\ $\sin^22\theta$, however, events down to 20 MeV
were used to increase the range of $E/L$, the ratio of the neutrino's energy
to its distance from the target to detection.  This was done because the plot
employed was intended to show the favored regions of $\Delta m^2$, and all
information about each event was used.  A likelihood analysis was utilized,
and the contours shown in Fig.~\ref{fig:1} are at 2.3 and 4.5 log-likelihood
units from the maximum.  If this were a Gaussian likelihood distribution,
which it is not (its integral being infinite), the contours would correspond
to 90\% and 99\% likelihood levels, but in addition they have been smeared
to account for possible systematic errors.  Those contours have been widely
misinterpreted as confidence levels---which they certainly are not---because
they were plotted along with confidence-level limits from other experiments.
This confusion of comparing likelihood levels for the LSND data with
confidence levels from other experiments is exacerbated by using the
20--36 MeV region
for the LSND data.  This higher background range makes some difference for
the 1993--5 data, but an appreciable difference for the parasitic 1996--7 runs
with an iron target,
which were at a low event rate, decreasing the ratio of signal/background
events.  This distorts the energy spectrum, making the higher $\Delta m^2$
values desirable for dark matter appear less likely.

The 1993--7 likelihood plot was compared at Neutrino `98 with
KARMEN\cite{ref:22} results
which used the new ``unified procedure''\cite{ref:new22} for confidence levels.
Because KARMEN saw no events, a limit based on that looked as if LSND were
ruled out.  KARMEN expected to see $2.88\pm0.13$ events, and a ``sensitivity''
contour corresponding to actually observing that event rate does not exclude
very much of the LSND parameter space.

\begin{figure}[tbph!]
\centerline{\epsfxsize=6cm
\epsfbox{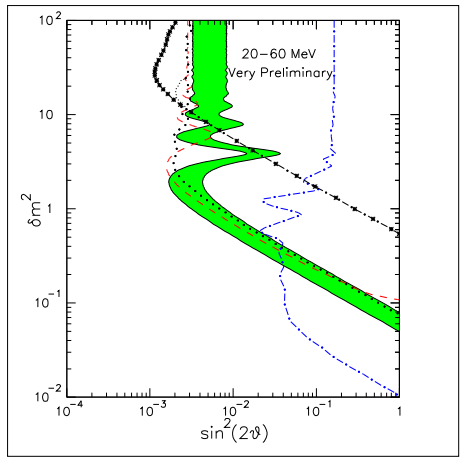}}
\caption{LSND upper and lower 90\% Bayesian confidence limits for
$\bar\nu_\mu\rightarrow \bar\nu_e$ oscillations using 1993-1997 data with
$20<E_e<60$ MeV.  Also shown is the 90\% Bayesian confidence upper limit
from KARMEN as of summer, 1998 (dashed), as well as upper limits from NOMAD
(dash-dot-star), E776 (dotted), and Bugey (dash-dot).}
\label{fig:2}
\end{figure}

\begin{figure}[tbph!]
\centerline{\epsfxsize=6cm
\epsfbox{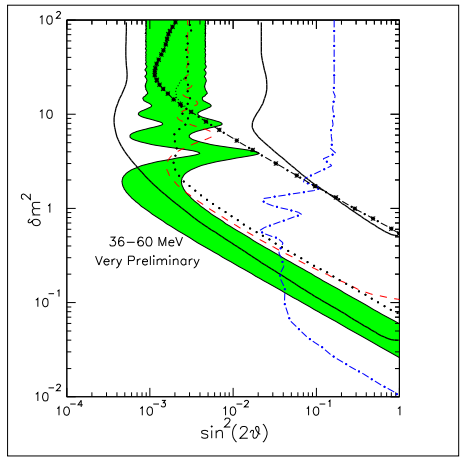}}
\caption{Same as Figure \ref{fig:2}, but from $36<E_e<60$ MeV data.
The added curves show the 80\% confidence band for the LSND $\nu_\mu\to\nu_e$
result.}
\label{fig:3}
\end{figure}

A fairer comparison of the two experiments is to use the same procedure for
each, so here Bayesian confidence levels are employed.  Because no attempt is
made to use $E/L$ to further constrain $\Delta m^2$, this is not the correct
way to determine favored regions of $\Delta m^2$.  The effect of excluding
the heavily contaminated 20--36 MeV region can be seen clearly from a
comparison of Figs.~\ref{fig:2} and \ref{fig:3}.  Figure~\ref{fig:2}, made
using data with $e^+$ energy between 20 and 60 MeV, seems to show that other
experiments exclude most of the LSND region, whereas Fig.~\ref{fig:3},
which uses the cleaner data with $e^+$ energy between 36 and 60 MeV, shows
that there is a wide range of $\Delta m^2$ not in contradiction with other
experiments.  Also shown in Fig.~\ref{fig:3} is the LSND $\nu_\mu\to\nu_e$
result\cite{ref:23} which although quite broad tends to favor higher
$\Delta m^2$ values.  This broadness results from the greater background in
this case, primarily because the observed process ($\nu_eC\to e^-X$) gives
only one signal instead of the two available in the $\bar\nu_\mu\to\bar\nu_e$
case.  While the fluctuation probability for $\nu_\mu\to\nu_e$ is only
$\sim10^{-2}$, the two ways of detecting oscillations are essentially
independent, providing some confirmation that a real effect is being observed.


\section{Number of Neutrino Types Needed for Dark Matter}
If neutrino dark matter were due to one neutrino, that would presumably be
$\nu_\tau$, and this would be ruled out if, as fits the Super-Kamiokande
data\cite{ref:19} best, the atmospheric anomalous $\nu_\mu/\nu_e$ ratio is
due to $\nu_\mu\to\nu_\tau$, since the mass-squared difference required is
$\Delta m^2_{\mu\tau}\sim10^{-3}$ eV$^2$, and the needed neutrino mass is
$94\Omega_\nu h^2\sim5$ eV (for 20\% neutrinos and $\Omega_m=1$, $h=0.5$ or
$\Omega_m=0.6$, $h=0.65$), where $h$ is the Hubble constant in units of
$\rm 100 km\cdot s^{-1}\cdot Mpc^{-1}$.  The only hope for $\nu_\tau$ dark
matter is if the atmospheric $\nu_\mu s$ are oscillating into sterile
neutrinos.  As discussed before, the large mixing angle required makes it
likely that the $\nu_s$ would provide a problem with
nucleosynthesis.\cite{ref:7}

A three-neutrino scheme could have $\nu_\mu\to\nu_\tau$ for the atmospheric
case, $\nu_\mu\to\nu_e$ (with $\Delta m^2_{\nu e}\ltwid10^{-5}$ eV$^2$) for
the solar $\nu_e$ deficit, and the three nearly mass degenerate neutrinos
could give the needed dark matter.  When this was first suggested\cite{ref:24}
there was a possible problem with neutrinoless double beta
decay.  While limits on that have improved, theoretical ways have been found
to ameliorate the problem.  If LSND is correct, however, this scheme is
certainly ruled out.

That leaves two-neutrino dark matter.  This scheme\cite{ref:24} requires
four neutrinos, with the solar deficit explained by $\nu_e\to\nu_s$ (and
both neutrinos quite light) the atmospheric effect due to
$\nu_\mu\to\nu_\tau$ (both of which are heavier and 
share the dark matter role) and the LSND $\nu_\mu\to\nu_e$ demonstrating
the mass difference between these two nearly mass-degenerate doublets.  Note
that the solar $\nu_e\to\nu_s$ is for the small mixing angle (or vacuum
oscillation), so $\nu_s$ does not affect nucleosynthesis.  The original
motivation for this mass pattern preceded LSND and was simply to provide some
hot dark matter, given the solar and atmospheric phenomena.  If LSND is
correct, it becomes the unique pattern.  However, just the $\nu_\mu\to\nu_\tau$
explanation of the atmospheric result alone forces two-neutrino dark matter.

This neutrino scheme was the basis for simulations\cite{ref:25} which showed
that two-neutrino dark matter fits observations better than the one-neutrino
variety.  The latter produces several problems at a distance scale of the
order of $10h^{-1}$ Mpc, particularly overproducing clusters of galaxies.
Whether the $\sim5$ eV of neutrino mass is in the form of one neutrino
species or two makes no difference at very large or very small scales, but
at $\sim10h^{-1}$ Mpc the larger free streaming length of $\sim5/2$ eV
neutrinos washes out density fluctuations and hence lowers the abundance of
galactic clusters.  In every aspect of simulations done subsequently the
two-neutrino dark matter has given the best results.  For example, a single
neutrino species, as well as low $\Omega_m$ models, overproduce void regions
between galaxies, whereas the two-neutrino model agrees well with
observations.\cite{ref:26}

\section{Supporting Information from Supernova Nucleosynthesis}
If LSND is correct, a sterile neutrino is required, and two-neutrino dark
matter is established.  A fourth light neutrino must not have the normal
weak interaction because of the measured width of the $Z^0$ boson.  Any
independent information favoring such a sterile
neutrino would support this four-neutrino scheme and the two-neutrino dark
matter.  Such information can come from that neutrino laboratory, the
supernova.

While $\Delta m^2_{e\mu}\sim6$ eV$^2$ is desirable for two-neutrino
dark matter, it apparently would cause a conflict with the
production of heavy elements in supernovae.  This $r$-process of rapid neutron
capture occurs in the outer neutrino-heated ejecta of Type II supernovae.  The
existence of this process would seem to place a limit on the mixing of
$\nu_\mu$ and $\nu_e$ because energetic $\nu_\mu\ (\langle E\rangle\approx25$
MeV) coming from deep in the supernova core could convert via an MSW transition
to $\nu_e$ inside the region of the $r$-process, producing $\nu_e$ of much
higher energy than the thermal $\nu_e\ (\langle E\rangle\approx11$ MeV).  The
latter, because of their charged-current interactions, emerge from farther out
in the supernova where it is cooler.  Since the cross section for $\nu_en\to
e^-p$ rises as the square of the energy, these converted energetic $\nu_e$
would deplete neutrons, stopping the $r$-process. 
Calculations\cite{ref:27} of this effect limit $\sin^22\theta$ for
$\nu_\mu\to\nu_e$ to $\ltwid10^{-4}$ for $\Delta m^2_{e\mu}\gtwid2$ eV$^2$, in
conflict with compatibility between the LSND result and a neutrino component of
dark matter.

The sterile neutrino, however, can not only solve this problem, but also rescue
the $r$-process itself.  While recent simulations have found the $r$-process
region to be insufficiently neutron rich,\cite{ref:28} very recent
realization of the full
effect of $\alpha$-particle formation has created a disaster for the
$r$-process.\cite{ref:29} The initial difficulty of too low entropy (i.e., too
few neutrons per seed nucleus, like iron) has now been drastically exacerbated
by calculations\cite{ref:29} of the sequence in which all available
protons swallow up neutrons to form $\alpha$ particles, following which
$\nu_en\to e^-p$ reactions create more protons, creating more $\alpha$
particles, and so on.  The depletion of neutrons by making $\alpha$ particles
and by $\nu_en\to e^-p$ rapidly shuts off the $r$-process, and essentially no
nuclei above $A=95$ are produced.

The sterile neutrino would produce two effects.\cite{ref:30}  First, there is
a zone, outside the neutrinosphere (where neutrinos can readily escape) but
inside the $\nu_\mu\to\nu_e$ MSW (``LSND") region, where the $\nu_\mu$
interaction potential goes to zero, so a $\nu_\mu\to\nu_s$ transition can occur
nearby, depleting the dangerous high-energy $\nu_\mu$ population.  Second,
because of this $\nu_\mu$ reduction, the dominant process in the MSW region
reverses, becoming $\nu_e\to\nu_\mu$, dropping the $\nu_e$ flux going into the
$r$-process region, hence reducing $\nu_en\to e^-p$ reactions and allowing the
region to be sufficiently neutron rich.

This description is simplified, since the atmospheric results show that the
$\nu_\mu$ and $\nu_\tau$ mix with a large angle, so wherever ``$\nu_\mu$''
is mentioned, this can equally well be ``$\nu_\tau$''.  In fact, if the
mixing is maximal and the $\nu_\mu$ and $\nu_\tau$ mix equally with the
$\nu_e$, one can show\cite{ref:30} that the $\nu_e$ flux above the second
resonance vanishes totally.  To keep the resonances separate and in the
proper order, they must occur below the weak freeze out radius, where the
weak interactions go out of equilibrium.  This requires a sufficiently large
$\Delta m^2_{e\mu(\tau)}$, and a value like 6 eV$^2$ satisfies this
requirement, enhancing the argument for hot dark matter.

\section{Supporting Information from Light-Element Nucleosynthesis}
Since the primordial baryon-to-photon ratio determined by the $^4$He
abundance\cite{ref:4} or by the deuterium to hydrogen ratio\cite{ref:5} is in
each case somewhat determined by the particular analysis of the data, it is not
clear at this time whether there is really any discrepancy remaining of
what was once an apparent crisis.  It is important, contrary to what was
once believed, that the sterile neutrino not produce a very big effect
on the $^4$He abundance.

The mechanism by which this might occur in the early universe is the
following.  If the potential is appropriate so that
$\bar\nu\to\bar\nu_s$ transitions occur instead of $\nu\to\nu_s$, such an MSW
transition could lead to a significant excess of $\nu_e$ over $\bar\nu_e$, so
that the $n/p$ ratio (and hence $^4$He) would be depleted prior to the
decoupling of the $\nu_en\to e^-p$ reaction.  The scarcity of initial sterile
neutrinos, which are produced only via mixing with active ones, makes the
dominant MSW transition active$\to$sterile and not the other way around.  The
small mass difference of the solar case makes $\bar\nu_e\to\bar\nu_s$ have a
negligible effect, but $\bar\nu_\mu\to\bar\nu_s$ and $\bar\nu_\tau\to\bar\nu_s$
with $\Delta m^2\sim6$ eV$^2$ could create a large lepton asymmetry which would
be transferred to $\nu_e$ via $\nu_\mu\to\nu_e$ and $\nu_\tau\to\nu_e$.
Calculations\cite{ref:31} show that for the four-neutrino model the effect
is small and could even resolve the remaining discrepancy, if any, but the
main point is that some other models with sterile neutrinos could produce
too big an effect.

\section{Conclusions}
A neutrino component of dark matter appears very probable, both from the
astrophysics and particle physics standpoints.  Despite the evidence for
$\Omega_m<1$, the one model which fits universe structure has $\Omega_m=1$,
with 20\% neutrinos and most of the rest as cold dark matter.  Open universe
and low-density models with a cosmological constant give extremely bad fits.
This conflict should be the source of future progress, but since there are
$10^2/\rm cm^3$ of neutrinos of each active species left over from the early
universe, the ultimate answer on neutrino dark matter will come from
determinations of neutrino mass.  While the solar and atmospheric evidences
for neutrino mass are important, the crucial issue is the much larger
mass-squared difference observed by the LSND experiment.  In the mass region
needed for dark matter, no other experiment excludes the LSND result, if
data from the different experiments are compared using the same procedures.

The resulting mass pattern, $\nu_e\to\nu_s$ for solar, $\nu_\mu\to\nu_\tau$
for atmospheric, and $\nu_\mu\to\nu_e$ for LSND, requires a sterile neutrino
and provides two-neutrino ($\nu_\mu$ and $\nu_\tau$) dark matter.  This
form of dark matter fits observational data better than the one-neutrino
variety.  Furthermore, the sterile neutrino appears to be necessary to
rescue the production of heavy elements by supernovae.  This particular mass
pattern does not cause any difficulty with the present near concordance in
primordial light element abundances, and it could even help with a remaining
small discrepancy.  In short, this four-neutrino pattern agrees with all
current neutrino mass information and hence makes more likely the existence
of hot dark matter.

\section*{Acknowledgments}
I wish to thank Steven Yellin for producing Figures 2 and 3.

\end{document}